\documentclass{aa}
\usepackage{graphicx,times}

\begin{document}
\title{The Soft X-ray Spectrum from NGC~1068 Observed with LETGS on Chandra}

\author{ A.C. Brinkman \inst{1}
         \and
         J.S. Kaastra \inst{1}
         \and
         R.L.J. van der Meer \inst{1}
         \and
         A. Kinkhabwala \inst{2}
         \and
         E. Behar \inst{2}
         \and
         S.M. Kahn  \inst{2}
         \and
         F.B.S. Paerels \inst{2}
         \and
         M. Sako \inst{3}
         }

\offprints{J.S. Kaastra}
\mail{J.Kaastra@sron.nl}

\institute{SRON, National Institute for Space Research,
             Sorbonnelaan 2, 3548 CA Utrecht, The Nether\-lands
             \and
           Columbia Astrophysics Laboratory, Columbia University, 550 West
	     120th Street, New York, NY 10027, USA
             \and
           Theoretical Astrophysics and Space Radiation Laboratory,
	     California Institute of Technology, MC 130-33, Pasadena, CA
	     91125, USA
           }

\date{Received  / Accepted}

\abstract{
Using the combined spectral and spatial resolving power of the Low
Energy Transmission Grating (LETGS) on board {\it Chandra}, we obtain separate
spectra from the bright central source of NGC~1068 (Primary region), and from a
fainter bright spot 4\arcsec\ to the {\small NE} (Secondary region).  Both
spectra are dominated by discrete line emission from H- and He-like ions of C
through S, and from Fe L-shell ions, but also include narrow radiative
recombination continua (RRC), indicating that most of the observed soft X-ray
emission arises in low-temperature ($kT_{\mathrm{e}} \sim$ few eV) photoionized
plasma.  We confirm the conclusions of Kinkhabwala et al.
(\cite{kinkhabwala02b}), based on {\it XMM-Newton} Reflection Grating
Spectrometer (RGS) observations, that the entire nuclear spectrum can be
explained by recombination/radiative cascade following photoionization, and
radiative decay following photoexcitation, with no evidence for the presence of
hot, collisionally ionized plasma.  In addition, we show that this same model
also provides an excellent fit to the spectrum of the Secondary region, albeit
with radial column densities roughly a factor of three lower, as would be
expected given its distance from the source of the ionizing continuum.  The
remarkable overlap  and kinematical agreement of the optical and X-ray line
emission, coupled with the need for a distribution of ionization parameter to
explain the X-ray spectra, collectively imply the presence of a distribution of
densities (over a few orders of magnitude) at each radius in the ionization
cone.  Relative abundances of all elements are consistent with Solar abundance,
except for N, which is 2--3 times Solar. Finally, the long wavelength spectrum
beyond 30~\AA\ is rich of L-shell transitions of Mg, Si, S, and Ar, and
M-shell transitions of Fe. The velocity dispersion decreases with increasing
ionization parameter, which has been deduced from the measured line intensities
of particularly these long wavelength lines in conjunction with the Fe-L
shell lines.
}

\titlerunning{The Soft X-ray Spectrum from NGC~1068}
\maketitle

\section{Introduction}

The bright Seyfert~2 galaxy, \object{NGC~1068}, has been studied extensively for
many years at optical, UV, IR, radio, and X-ray wavelengths.  In the unified
model of active galactic nuclei (AGN) (Miller \& Antonucci \cite{miller};
Antonucci \& Miller \cite{antonucci85}; Antonucci \cite{antonucci93}), the soft
X-ray spectra of Seyfert~2 galaxies are expected to be strongly affected by
emission and scattering in a medium irradiated by the nuclear continuum.
However, soft X-ray emission can also be produced by collisionally-heated gas
associated with shocks in starburst regions (Wilson et al.  \cite{wilson}).
High resolution X-ray spectroscopy, now becoming available with the grating
spectrometers on {\it Chandra} and {\it XMM-Newton}, can provide a means of
distinguishing between these two interpretations.

Recently, Kinkhabwala et al.  (\cite{kinkhabwala02b}) presented the X-ray
spectrum of the nuclear regions in NGC~1068 obtained with the Reflection Grating
Spectrometer (RGS) on {\it XMM-Newton}.  They show that the soft X-ray emission
is dominated by lines from H-like and He-like ions of all low-Z elements from C
through Si, as well as a complex of L-shell lines from Ne-like through Li-like
Fe.  Most of the lines are significantly blueshifted and broadened, with
characteristic velocities $\sim$ several hundred km~s$^{-1}$.  Of particular
interest is the presence of strong and narrow RRC of C, N, O, and Ne ions, which
imply that most of the observed soft X-ray flux arises in low-temperature ($kT
\sim$ few eV) plasma.  There is also excess emission (relative to pure
recombination) in all higher series resonance lines of the H-like and He-like
species, which is shown to be a result of direct photoexcitation by the
continuum.

Kinkhabwala et al.  (\cite{kinkhabwala02b}) further present a new quantitative
model for the nuclear regions of NGC~1068 that provides an excellent fit to
their observed spectrum.  The model involves a cone of warm plasma photoionized
and photoexcited by the same nuclear power-law continuum (Kinkhabwala et al.
\cite{kinkhabwala02a}).  In this model, each ionic line series is fit
independently, with only three free parameters:  the radial column density for
that ion, the velocity width, and the overall normalization (the product of the
nuclear luminosity and the covering factor).  Interestingly, the derived column
densities are consistent with those measured using absorption features in high
resolution X-ray spectra of Seyfert~1 galaxies, thereby providing strong support
for the unified model.

Given the 15\arcsec\ spatial resolution of the {\it XMM-Newton} telescopes, it
was not possible with the RGS data to derive any useful information on the
spatial dependence of the observed emission line spectrum.  This issue is best
addressed with the transmission grating spectrometers on the {\it Chandra} X-ray
Observatory.  In this paper, we present the data acquired with the LETGS (see
also the companion paper presenting HETGS data by Ogle et al.  \cite{ogle}).
Earlier {\it Chandra} imaging observations of NGC~1068 (Young et al.
\cite{young}) had shown the presence of a bright compact source, $\sim$
1\arcsec.5 or $108$~pc in extent (1\arcsec = $72$~pc), taking the distance to
NGC~1068 to be 14.4~Mpc (Bland-Hawthorn et al.  \cite{bland-hawthorn}).  This is
coincident with the brightest radio and optical emission, together with an
extended cone of emission (out to 7\arcsec; 504~pc) pointing to the NE, which
coincides with the NE radio lobe and gas in the narrow line region.  A discrete
region of emission is observed in the cone at about 4\arcsec\ ($288$~pc) from
the bright compact source.  We show here that the spectrum of this region
differs significantly from that of the compact central source, but that it is
also well described by the same radiation-driven model presented by Kinkhabwala
et al.  (\cite{kinkhabwala02b}), albeit with lower radial column densities.  The
lower column densities are expected, and are a natural consequence of the larger
distance of this extended gas from the source of the ionizing continuum.

The LETGS also has a significantly larger bandpass than the RGS and allows us to
investigate the nature of the spectrum down to the region of the Fe K-shell
transitions near $2$~\AA, and out beyond $35$~\AA, where we detect L-shell
transitions of Ar, S, Si, and Mg.

\begin{figure}
\resizebox{\hsize}{!}{\includegraphics[angle=0]{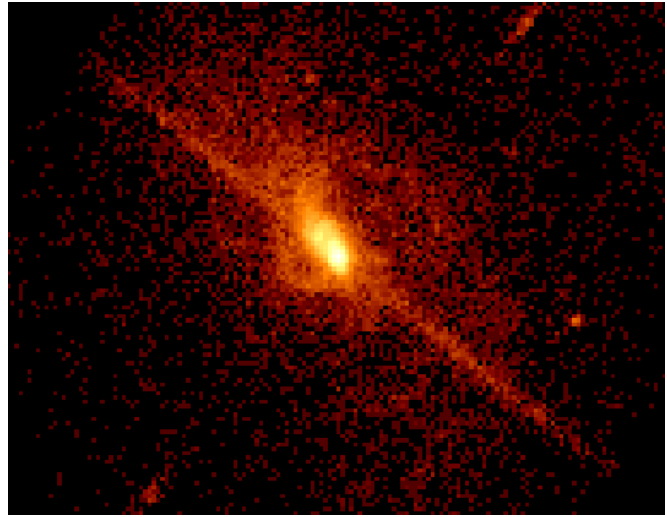}}
\caption{ LETGS zero order image.  North is up and East is left.  The strong bar
(top left-bottom right) represents "out of time events" from the CCD-readout.
The beginning of the spectral image, minus order bottom left and positive
orders, top right can just be seen.  Notice the dispersion direction is close to
perpendicular to the extended NE-direction.  The point source in the lower right
quadrant has a very hard spectrum.  The angular size of the total
image is 60\arcsec $\times$ 75\arcsec.  }
\label{fig:zeroimage}
\end{figure}

\begin{figure}[!hb]
\resizebox{\hsize}{!}{\includegraphics[angle=0]{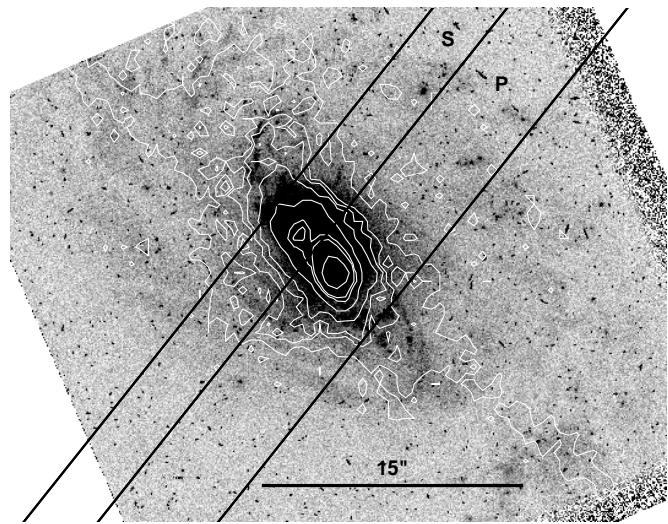}}
\caption{Zero-order X-ray intensity contours plotted over an optical
[\ion{O}{iii}] (5007 \AA) HST image (Bruhweiler et al. \cite{bru}).  The
Primary (``P'') and Secondary (``S'') extraction regions are indicated.
The X-ray/optical correlation is remarkably good. }
\label{fig:oxygen}
\end{figure}

\section{Observation and Data Analysis}

Since the extended nature and the orientation of the X-ray emission from
NGC~1068 were known prior to the time the grating observations were planned, the
roll angle was selected such that the dispersion angle was nearly perpendicular
to the NE extent, making it possible to separately study the emission from the
central nuclear component (hereafter, the Primary region), and a bright spot in
the extended ionized cone (hereafter, the Secondary region).  The HETGS (Ogle et
al.  \cite{ogle}), and the LETGS (this work) observed the source on December 4
and 5, 2000, for 46 ks and 77 ks, respectively.  The LETGS configuration for
this measurement involved the ACIS-S as the readout detector, in place of the
usual HRC.  We used CIAO version 2.1 to generate the event file.  In order to
verify the wavelength scale as generated with CIAO, we ran a Capella calibration
observation (OBSID~00055), and did a recursion analysis based on a few strong
spectral lines.  There appears to be a very small systematic offset proportional
to the wavelength, which is close to a similar systematic offset found during
the analysis of the HETGS measurement of NGC~4151 (Ogle et al.
\cite{ogle2000}).  The correction is $+0.7$~m\AA\ per \AA.

The LETGS zero order image is displayed in Fig.~\ref{fig:zeroimage}.  The
diagonal streak in this image is {\em not} the dispersed spectrum, it arises
from ``out of time" events in the CCD-readout.  The short wavelength edges of
the spectrum can be discerned as spots in the lower left and upper right of the
figure.  Note that the source extent is nearly perpendicular to the dispersion
direction.  In Fig.~\ref{fig:oxygen}, this X-ray image is shown as a contour
plot overlayed on an [\ion{O}{iii}] optical image obtained by HST.  As can be
seen, the X-ray extension correlates well with the extended [\ion{O}{iii}]
emission.  We have indicated in Fig.~\ref{fig:oxygen} the source extraction
regions, which we use to isolate the spectrum of the Primary and Secondary
regions.

\begin{figure}
\resizebox{\hsize}{!}{\includegraphics[angle=-90]{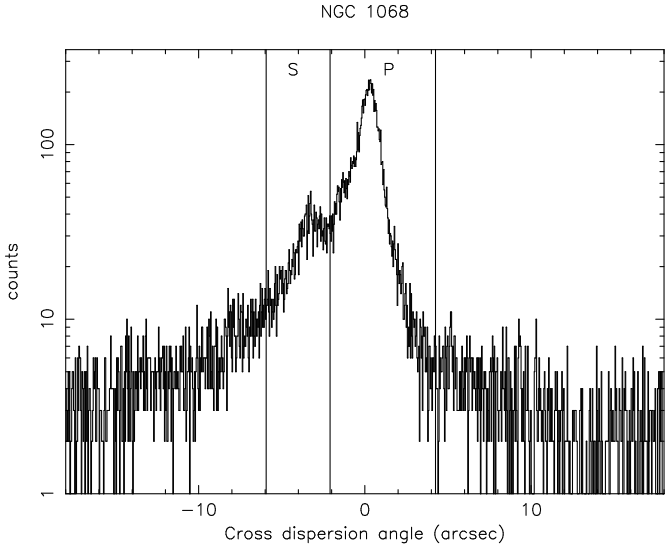}}
\caption{Intensity as a function of the cross dispersion direction.  The
Primary (``P'') and Secondary (``S'') regions are indicated.}
\label{fig:crossdispersion}
\end{figure}

A spatial cut through this zero order image in the cross-dispersion direction is
plotted in Fig.~\ref{fig:crossdispersion}.  Clearly visible on the left side
(NE-direction) is the small Secondary peak, which corresponds to our Secondary
region.  Here again, the extraction regions for the Primary and Secondary
spectra are indicated.  The boundary between them was chosen based on the
\ion{O}{vii} line ratio variation along the cross-dispersion direction (see
Sect.~3).

The zero order profile in the dispersion direction of the Primary region is
well represented by a Gaussian with the centroid at $-0.0056\pm0.0007$~\AA,
and with a width given by $\sigma = 0.0317\pm0.0008$~\AA.  This profile is
slightly broader than the instrument profile for a point source
($\mathrm{FWHM} = 0.06$~\AA; $\sigma = 0.0255$~\AA), and is therefore
affected by the geometry of the source extent.  Similarly, the profile of the
Secondary region in the dispersion direction is represented by a Gaussian,
with the centroid at $+0.032\pm0.002$~\AA, and with a width
$\sigma = 0.062\pm0.002$~\AA.

We checked for time variability in the zero order flux.  As is to be expected
for this spatially extended source, we find no significant variations on
timescales in the range $\sim$ few seconds up to the length of the observation.

\begin{figure*}
\resizebox{\hsize}{!}{\includegraphics[angle=0]{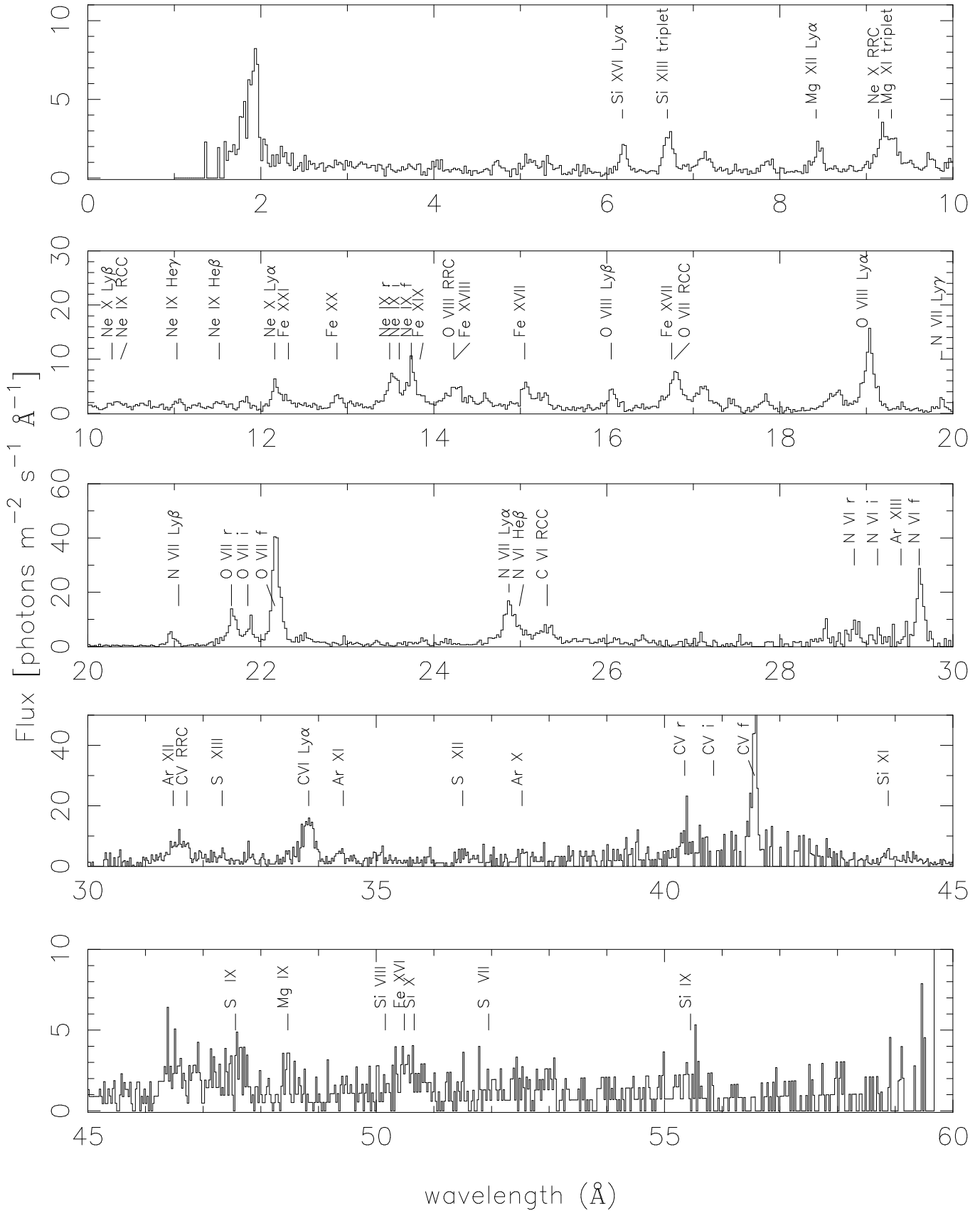}}
\caption{Spectrum of the Primary region between 1.5 and $60$~\AA, in five panels
to cover the full range.  The bin size is equal to $0.025$~\AA.~ Indicated are
some of the strongest lines.  All panels are the positive and negative orders
taken together.}
\label{fig:Primary_spec_added}
\end{figure*}

\begin{figure*}
\resizebox{\hsize}{!}{\includegraphics[angle=0]{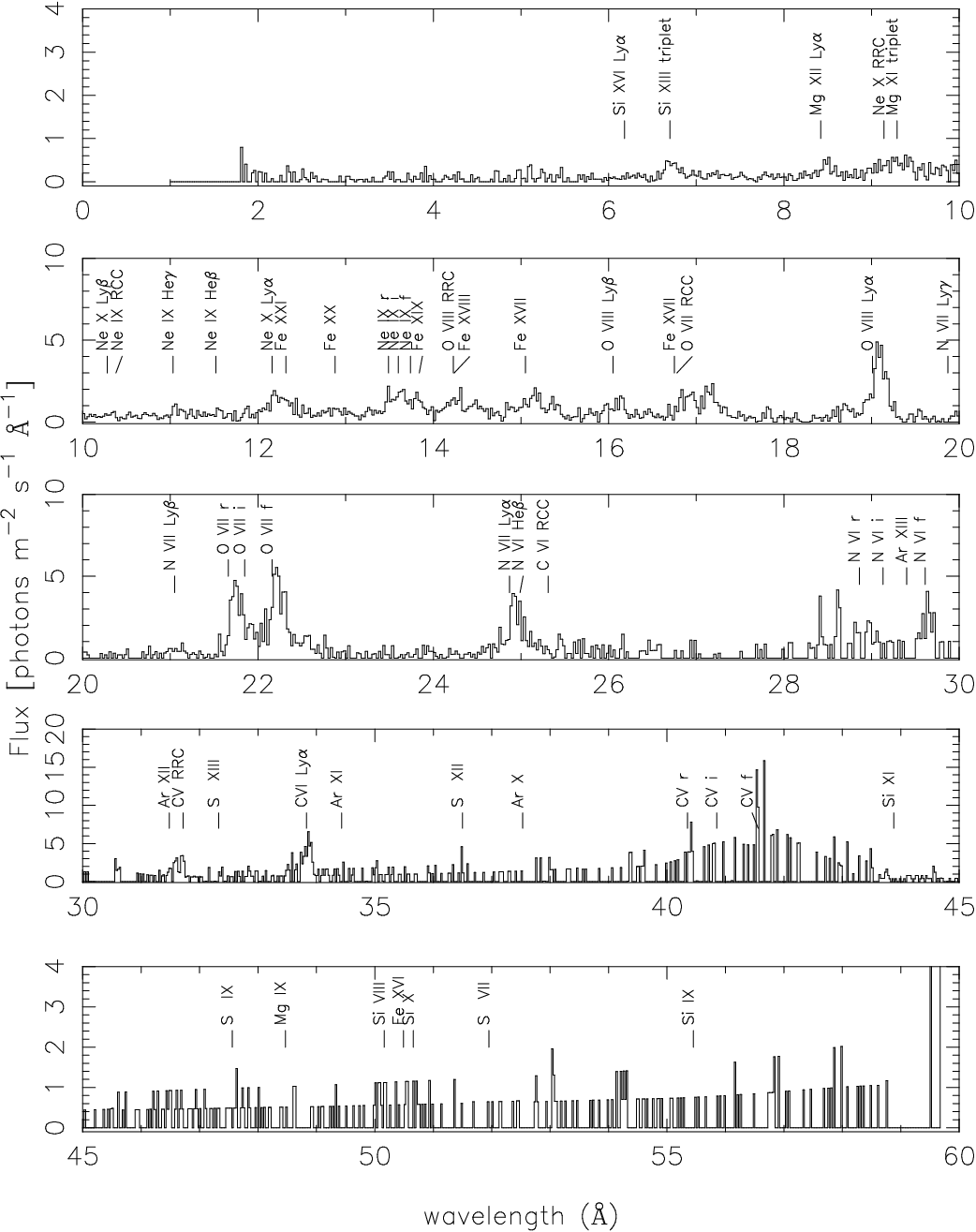}}
\caption{Spectrum of the Secondary region, wavelength scales as in previous figure for easy comparison.}
\label{fig:Secondary_spec_added}
\end{figure*}

\section{Spectral Analysis}

We present spectra of the Primary and Secondary regions in
Figs.~\ref{fig:Primary_spec_added} and \ref{fig:Secondary_spec_added},
respectively.  We combine $m=-1$ and $m=+1$ orders to obtain these spectra.
Both spectra are plotted over the full range of the LETGS of
$\sim 1.5-60$~\AA.

The Primary region contains approximately ninety percent of the flux ("P" region
in Figs.~\ref{fig:oxygen} and \ref{fig:crossdispersion}).  As is expected, this
spectrum closely resembles the RGS spectrum (Kinkhabwala et~al.
\cite{kinkhabwala02b}), which contains all of the soft-X-ray flux from NGC~1068.
In addition to the H-like and He-like series of C, N, O, Ne, Mg, and Si and Fe
L-shell transitions (from \ion{Fe}{xvii}--\ion{Fe}{xxiv}), which were observed
by the RGS, the LETGS also extends down to Fe~K at $1.8$~\AA, and, at longer
wavelengths, from the He-like C triplet at $40-42$~\AA\ (which is just outside
the RGS range) out to $60$~\AA, resolving a forest of lines from L-shell
transitions in mid-Z ions such as Ar, S, Si, and Mg.  We also note that lines in
the $\sim 5-10$~\AA\ region of the specrum, though detected by the RGS, are
better resolved by the LETGS.

As was also found in the RGS spectrum, the LETGS resolves several prominent RRC,
which, in addition to the relative strength of the forbidden line in the He-like
triplets, provides the main evidence for recombination in low temperature,
photoionized plasma.  Many higher order transition lines, observed by the RGS,
are also resolved by the LETGS; these lines are indicative of photoexcitation
(e.~g., Sako et~al.  \cite{sako}).

Overall, many features observed in the Primary spectrum are also observed in the
spectrum of the Secondary region ("S" region in Figs.~\ref{fig:oxygen} and
\ref{fig:crossdispersion}), which comprises approximately ten percent of the
total flux from NGC~1068.

\begin{table}
\caption{He-like $R$ ratios for Primary and Secondary region spectra.}
\label{tab:Rratios}
\centering
\begin{tabular}{|l|cc|}
\hline
Ion     & Primary & Secondary\\
\hline
\ion{C}{v}   &  $11\pm 17$   &   $1.5\pm 1.1$  \\
\ion{N}{vi}  &  $3.3\pm 1.5$ &   ---       \\
\ion{O}{vii} &  $5.4\pm 0.9$ &   $4.5\pm 2.0$ \\
\ion{Ne}{ix} &  $2.3\pm 0.5$  &  $1.4\pm 0.6$ \\
\hline
\end{tabular}
\end{table}

\begin{table}
\caption{He-like $G$ ratios for Primary and Secondary region spectra.}
\label{tab:Gratios}
\centering
\begin{tabular}{|l|cc|}
\hline
Ion     & Primary & Secondary\\
\hline
\ion{C}{v}   & $2.5\pm 1.4$     & $1.8\pm 1.3$    \\
\ion{N}{vi}  & $3.0\pm 1.1$     & $1.1\pm 0.9$    \\
\ion{O}{vii} & $4.6\pm 0.9$     & $1.6\pm 0.4$    \\
\ion{Ne}{ix} & $2.0\pm 0.3$     & $1.4\pm 0.4$    \\
\hline
\end{tabular}
\end{table}

One main difference between the two spectra is in the He-like triplet ratios,
$R=f/i$ and $G=(i+f)/r$, which are listed for several He-like ions in the
Primary and Secondary spectra in Tables~\ref{tab:Rratios} and \ref{tab:Gratios}.

The $G$ ratios in the Primary spectrum are inconsistent with either pure
recombination or collisional ionization equilibrium, suggesting the presence of
photoexcitation, as was concluded for the RGS spectrum (Kinkhabwala et al.
\cite{kinkhabwala02b}).  The $G$ ratios in the Secondary region are consistent
with hot collisional plasma or with photoexcitation in a recombining,
photoionized plasma.

The $R$ ratios are overall less useful for distinguishing between pure
recombination and collisional ionization equilibrium.  They are also more
difficult to measure due to the weakness and/or blending of the intercombination
line in most triplets.  However, we note that the well-measured $R$ ratio for
\ion{O}{vii} is significantly greater than expected for pure recombination, as
was also found in Kinkhabwala et al.  (\cite{kinkhabwala02b}), who suggested
that this may be due to inner-shell ionization of \ion{O}{vi} (e.g., Kinkhabwala
et al.  \cite{kinkhabwala02a}).

\begin{figure}
\resizebox{\hsize}{!}{\includegraphics[angle=-90]{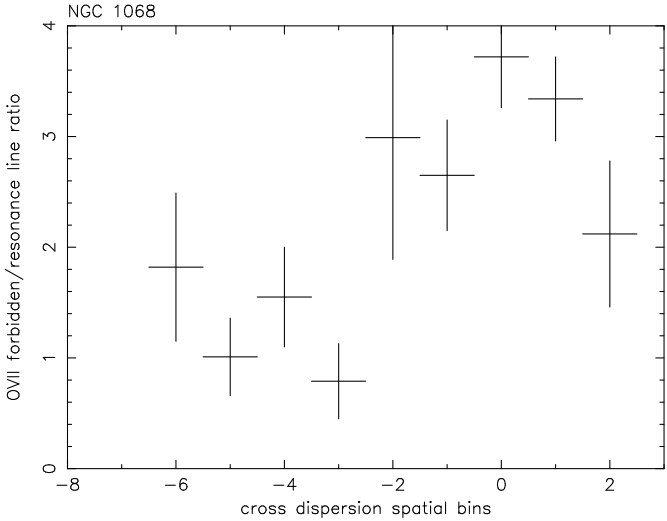}}
\caption{Forbidden line to resonance line ratio of the \ion{O}{vii} triplet as a
function of cross dispersion angle.  The bins along the cross dispersion axis
are one arcsec wide; negative numbers are in the NE direction, positive numbers
in the SW direction and bin zero is centered at the maximum intensity.  One
clearly notices an abrupt change over from a value of 3.6$\pm$0.6 for the
Primary region to 1.0$\pm$0.3 for the Secondary region, around the
boundary between these regions at $-$2\arcsec.  }
\label{fig:oxycross}
\end{figure}

Since the \ion{O}{vii} triplet is the strongest feature in both spectra and
since the line ratios are different for both spectra, we measured the
forbidden-to-resonance line ratio $f/r$ as a function of cross-dispersion, in
steps of 1\arcsec, see Fig.~\ref{fig:oxycross}.  There are clearly two distinct
regimes, the Primary region with a high ratio of $f/r=3.6\pm0.6$, and the
Secondary region with a value of $1.0\pm 0.3$.  This changeover occurs within a
spatial extent of less than one arcsec.

\subsection{Fits Using an Irradiated Cone Model}

We find that the H-like and He-like ionic line series observed in the Primary
and Secondary region spectra are fit well with a simple irradiated cone model,
as was used for the RGS spectrum of NGC~1068 (Kinkhabwala et~al.
\cite{kinkhabwala02b}).  The cone is irradiated at its tip by an inferred
nuclear power-law continuum.  Though our view to the nucleus is completely
obscured by intervening material (dusty torus), our perpendicular view of the
irradiated cone is unobscured, allowing for observation of
recombination/radiative cascade following photoionization, and radiative decay
following photoexcitation.  This simple irradiated cone model has been
incorporated into XSPEC as the local model {\it photo} (Kinkhabwala et~al.
\cite{kinkhabwala02a}), which has been used to generate the fits to the Primary
and Secondary regions as presented below.

We assume that the overall normalization, $fL_X$, which is the covering factor
$f=\Omega/4\pi$ times the power-law luminosity $L_X$ is the same for all ions.
For ease of comparison of the LETGS results with the RGS results and between the
Primary and Secondary region spectra, we assume the same value for the
normalization of $fL_X=10^{43}$~ergs~s$^{-1}$.

We also assume a single radial velocity distribution specified by Gaussian width
$\sigma_v^{\mathrm{rad}}$.  The velocity width needed to explain the Primary
region spectrum of $\sigma_v^{\mathrm{rad}}=100$~km~s$^{-1}$, however, is a
factor of two smaller than that required to explain the RGS spectra.  For
simplicity, we use this velocity width for the Secondary region as well.

The only free parameters left are the individual radial ionic column densities.
We allow for individual line absorption, but assume that photoelectric
absorption is negligible.  With this assumption, we can simplify the
calculations by irradiating each ionic column density separately with an
initially unabsorbed power law, since the effect of individual line absorption
due to all other ions, which removes only a small fraction of the flux, can be
safely neglected.  Alternatively, using the method employed by {\it photo}, we
can irradiate all ionic column densities together, assuming all ions have the
same radial distribution.  The radial ionic column densities we infer below are
indeed consistent with the assumption of negligible photoelectric absorption,
and we have checked that either method yields the same result.  Moreover, as we
argue in Sect.~\ref{sec:discussion}, there likely exists a broad range of densities
at each radius, implying that the assumption that all ions have a similar radial
distribution may also be physically correct.

We demonstrate below that these simple model assumptions allow for a good fit to
both the Primary and Secondary region spectra.

\subsubsection{Primary Region}

\begin{figure*}[!ht]
\resizebox{\hsize}{!}{\includegraphics[angle=+90]{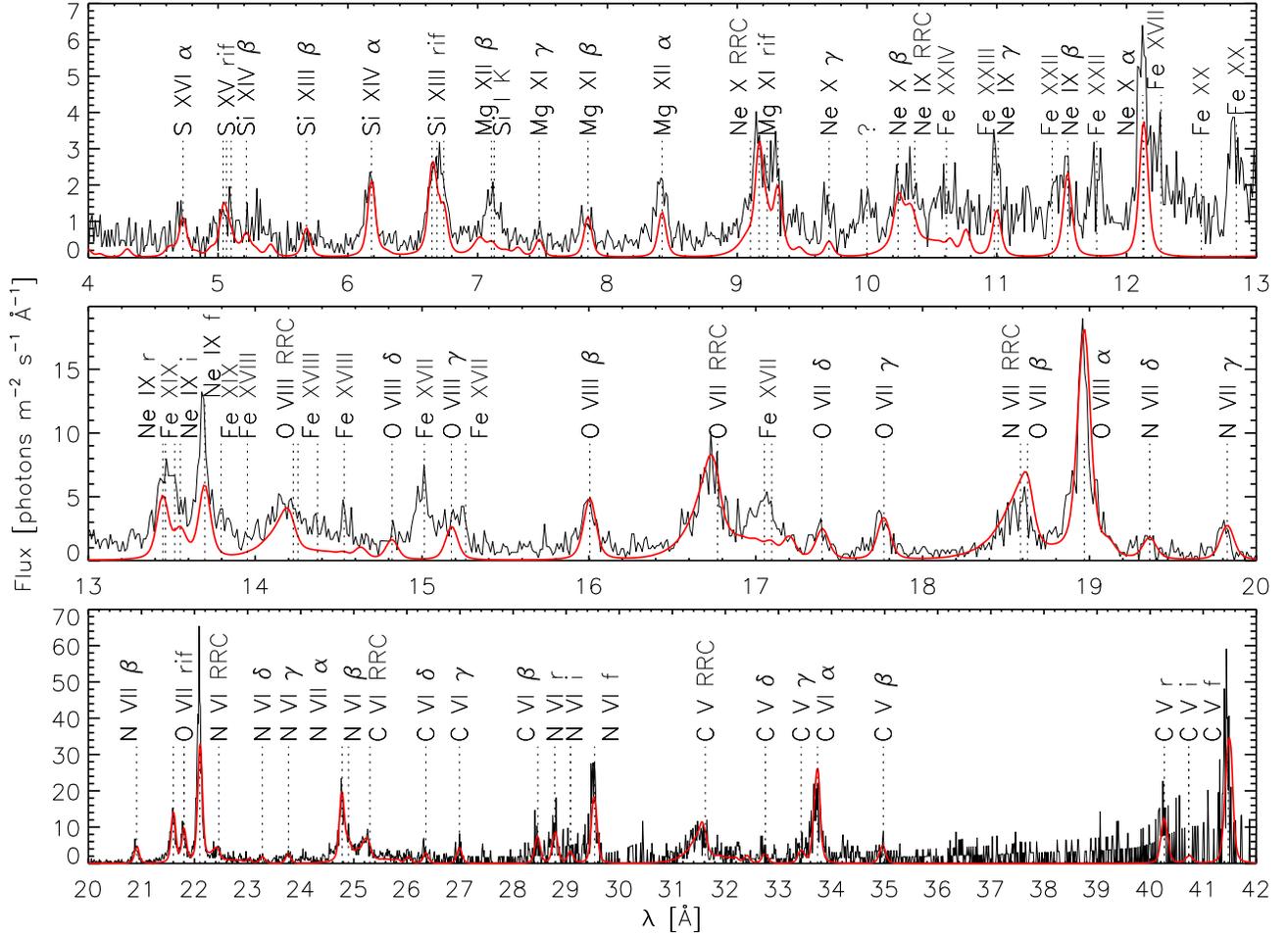}}
\caption{Fluxed spectrum of the first order of the Primary region, together with
the model.  The model does not yet include Fe L-shell transitions.}
\label{fig:Primary}
\end{figure*}

Using the above-quoted global values of $fL_X=10^{43}$~ergs~s$^{-1}$ and
$\sigma_v^{\mathrm{rad}}=100$~km~s$^{-1}$, we have fit the individual radial
ionic column densities.  The best fit values are given in
Table~\ref{tab:densities}, and the final fit using these values is shown in
Fig.~\ref{fig:Primary}.  These column densities are very similar to the results
obtained for the {\small RGS} spectrum.

The ratio of the ground state {\small RRC} to the 2p$\rightarrow$1s transitions
for each ion rules out the presence of any additional hot, collisional plasma.
The forbidden lines in the He-like triplets of N, O, and Ne, however, appear
significantly stronger than predicted.  This is likely due to inner shell
ionizations in the corresponding Li-like ions, which enhance only the forbidden
line in the triplet (for further explanation, see, e.~g., Kinkhabwala et~al.
\cite{kinkhabwala02a}).  This observation and proposed mechanism was also noted
in the {\small RGS} paper (Kinkhabwala et~al.  \cite{kinkhabwala02b}).

\subsubsection{Secondary Region}

\begin{figure*}[!ht]
\resizebox{\hsize}{!}{\includegraphics[angle=+90]{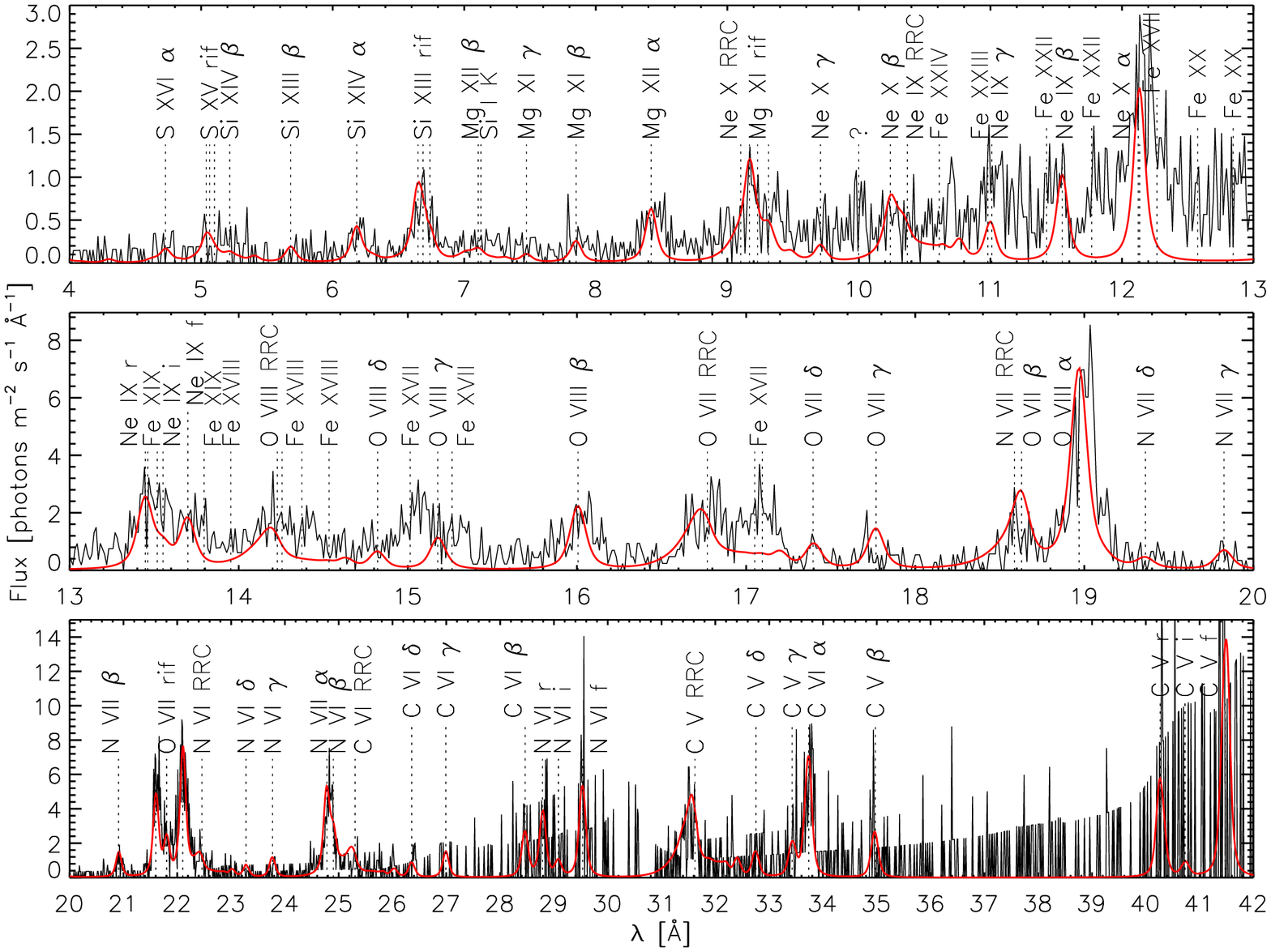}}
\caption{Spectrum of the Secondary region, together with the model.}
\label{fig:Secondary}
\end{figure*}

Using the same global values as above, we have fit the column densities for the
Secondary region.  The best fit values are given in Table~\ref{tab:densities},
and the final fit is shown in Fig.~\ref{fig:Secondary}.  The column densities
for the Secondary region are a factor of a few less than the column densities
inferred for the Primary region.

The explanation for this enhancement of the resonance line for lower column
density is simple (Kinkhabwala et~al.  \cite{kinkhabwala02b}).  Keeping
$\sigma_v^{\mathrm{rad}}$ constant, at low column densities the resonance line
is dominant in the He-like triplet, since this transition has a high oscillator
strength and is unsaturated.  At higher column densities this transition begins
to saturate, and recombination begins to increase in significance (since the
photoelectric edge saturates only at high column density).  Generically, then,
at low column densities, the resonance line is relatively strong, but at high
column densities, it saturates, becoming relatively weaker compared to lines due
to pure recombination.

The ratio of the {\small RRC} to the 2p$\rightarrow$1s transitions again rules
out the presence of a significant additional hot, collisional plasma.  Though,
given the lower quality of the Secondary spectrum compared to the Primary
spectrum, this conclusion is correspondingly less robust.  However, the presence
of strong higher order lines (np$\rightarrow$1s), such as \ion{N}{vii}
Ly$\gamma$ and \ion{O}{vii} He$\delta$, which can only be explained by
photoexcitation, strengthens the argument against an additional hot, collisional
plasma contribution (Kinkhabwala et~al.  \cite{kinkhabwala02b}).
The formal upper limit to the emission measure of any hot, collisional
plasma in the 0.2--1~keV temperature range is $10^{62}$~cm$^{-3}$ for the
Secondary region and $2\times 10^{62}$~cm$^{-3}$ for the Primary region.

We also note that the temperatures for the {\small RRC} are the same as for the
Primary region (in particular, compare the {\small RRC} of \ion{O}{vii} in both
spectra).

\begin{table}
\caption{NGC~1068 Ionic column densities of Primary and Secondary regions.
Italicized values for the temperatures were taken arbitrarily for ions which do
not show clear {\small RRC}.}
\label{tab:densities}
\centering
\begin{tabular}{|llll|}
\hline
Ion     & N$_{\rm ion}$ Primary & N$_{\rm ion}$ Secondary &   $kT$ (eV) \\
\hline
\ion{C}{v}    &   7.$10^{17}$   &     3.$10^{17}$   & 2.5\\
\ion{C}{vi}   &   8.$10^{17}$   &     4.$10^{17}$   & 4   \\
\ion{N}{vi}   &  3.$10^{17}$    &     1.$10^{17}$   & 3  	    \\
\ion{N}{vii}  &  5.$10^{17}$    &     1.$10^{17}$   & 4  \\
\ion{O}{vii}  &  7.$10^{17}$    &     2.$10^{17}$   & 4  \\
\ion{O}{viii} &  6.$10^{17}$    &     2.5.$10^{17}$ & 6  \\
\ion{Ne}{ix}  &  2.$10^{17}$    &     8.$10^{16}$   & 10  \\
\ion{Ne}{x}   &  1.$10^{17}$    &     6.$10^{16}$   & 10  \\
\ion{Mg}{xi}  &  1.$10^{17}$    &     2.5.$10^{16}$ & {\it 10} \\
\ion{Mg}{xii} &  1.$10^{17}$    &     2.$10^{16}$   & {\it 10}  \\
\ion{Si}{xiii}&  1.1.$10^{17}$  &     3.$10^{16}$   & {\it 10}  \\
\ion{Si}{xiv} &   1.1.$10^{17}$ &     2.$10^{16}$   & {\it 10}  \\
\ion{S}{xv}   &    4.$10^{16}$  &     1.$10^{16}$   & {\it 10}  \\
\ion{S}{xvi}  &    6.$10^{16}$  &     1.$10^{16}$   & {\it 10}   \\
\hline
\end{tabular}
\end{table}

\subsection{Elemental Abundance Estimates}

\begin{figure}
\resizebox{\hsize}{!}{\includegraphics[angle=-90]{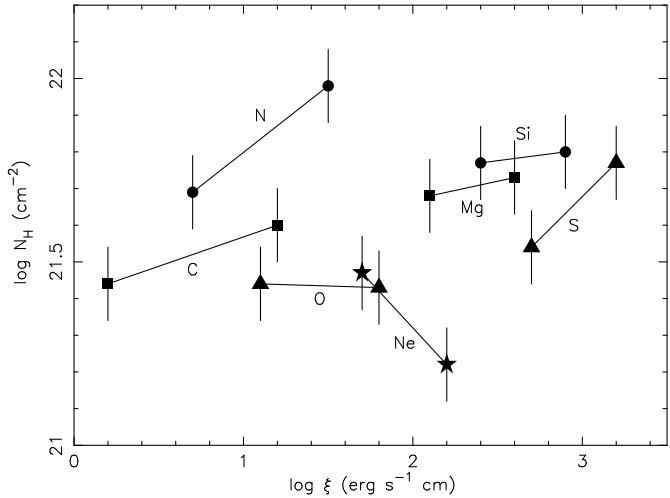}}
\caption{Derived hydrogen column density as a function of ionization parameter
for C, N, O, Ne, Mg, Si and S.}
\label{fig:psi.1068}
\end{figure}

In order to obtain an estimate of the relative elemental abundances, we derive a
total hydrogen column density from the individual ionic column densities
assuming Solar abundances.  This hydrogen column density $N_{\rm H}$ is a
function of the ionization parameter ($\xi$), according to the $\xi$ in which
each ion forms.  For this analysis, we employ the same set of photoionization
balance calculations from {\small XSTAR} (Kallman \& Krolik \cite{kallman}) as
used in the analysis of the Seyfert~1 galaxy NGC~5548 (Kaastra et al.
\cite{kaastra2002}) and assume that the observed flux for each ion originates
predominantly from plasma at an ionization parameter where the relative
concentration of that ion reaches a maximum.  Adopting Solar abundances from
Anders \& Grevesse (\cite{anders}), except for the recently improved measurement
of the Solar O abundance taken from Allende Prieto et al.  (\cite{allende}), we
obtain $N_{\rm H} (\xi)$ for the Primary component as plotted in
Fig.~\ref{fig:psi.1068}.

Note that these column densities are consistent with the estimated column of log
$N_{\rm H}$~=~$\sim$21.9 derived from the L-shell lines (Sect.~\ref{sect:L-lines}).
As explained in detail in that section, this value of $N_{\rm H}$ is consistent
with the observed line fluxes measured for the Fe lines assuming that they arise
exclusively from photoexcitation.  However, since most of the bright Fe lines
are saturated, this estimate is very uncertain and thus of less value for
abundance measurements.  We estimate the accuracy of the column densities based
on the K-shell lines to be within $\sim$~25~\% (relative to each other).  The
error bars in Fig.~\ref{fig:psi.1068} reflect these

The data point corresponding to H-like Ne in Fig.~\ref{fig:psi.1068}
seems to be somewhat low compared with the general trend suggested by the other
ions.  Therefore we consider this point on the plot less trustworthy for an
abundance analysis.  This inconsistency could be due to the Ly$\alpha$ line of
\ion{Ne}{x} being contaminated by \ion{Fe}{xvii}.

Deviations from Solar abundances in Fig.~\ref{fig:psi.1068} would appear as
clear vertical offsets between elements that cover similar regions of $\xi$.
However, it can be seen that obvious offsets between elements are not seen.  In
fact, the column density $N_{\rm H}$ does not vary by much in the observable
range of $\xi$.  This implies that for the most part the {\it relative}
abundances (no measurement is available for H) are consistent with Solar, or at
least that there are no conclusive deviations, of more than about $\pm$50\%,
from Solar abundances.  On the face of it, this is in stark contrast with
previous moderate-resolution X-ray observations of NGC~1068 as well as with UV
and optical measurements from the narrow line region.  Netzer \& Turner
(\cite{netzer}) as well as Marshall et al.  (\cite{marshall}) have fitted the
ASCA and BBXRT X-ray spectra, respectively, of NGC~1068 with models that require
an Fe/O abundance ratio of more than 8 times Solar.  This could be attributed to
the severely incomplete Fe-L model that was used in those works.  Also,
photoexcitation was not treated explicitly as shown to be necessary in
Kinkhabwala et al.  (\cite{kinkhabwala02b}).  In the ASCA case, the weak signal
in the low-energy region of the O features might also be to blame.  On the other
hand, the result of those papers that all other abundances except for Fe and O
are Solar is consistent with the present measurement, with the exception of N as
discussed below.  Nitrogen apparently was not included in those earlier models.

According to Fig.~\ref{fig:psi.1068}, N is the only element that seems to lie
consistently above the data points of other elements (namely O and C) with which
it overlaps in $\xi$ space.  The N abundance enhancement found here is of the
order of a factor 2$-$3 compared to C and O (also observed in Kinkhabwala et al.
\cite{kinkhabwala02b}).  This could be due to an enrichment of the AGN medium by
WR type stars close to the nucleus.  An enrichment of nitrogen has also been
found in other AGN, for example in IRAS 13349+2438 (Sako et al.
\cite{sako2001}).  Finally, we note that for the Secondary region, we find very
similar results only that the $N_{\rm H}$ is about $10^{21}$~cm$^{-2}$, which is
3$-$5 times smaller than for the Primary region.  Additionally, the error bars
for the Secondary region are obviously larger than for the Primary region.

\section{Line Broadening and Line Shifts\label{sect:line}}

\subsection{Line Broadening}

\begin{table*}
\caption{Line strengths and velocity shifts of some of the stronger lines in the
Primary and Secondary region.  Line strengths are at the source and in units of
$10^{48}$~photons~s$^{-1}$.  We adopt a
galactic absorption $N_{\mathrm{H}}=2.95\times 10^{20}$~cm$^{-2}$ (Murphy et al.
\cite{mur}) and a redshift of $z=0.00379\pm0.00001$ (Huchra et al.
\cite{huchra}).  The errors quoted are statistical errors; the systematic error
in the velocities in columns 5 and 6 range from 100 to 30~km~s$^{-1}$ for Si and
O, respectively.  All velocity values are quoted with respect to the systemic
velocity.  The line strengths and $\Delta v$ values of the \ion{Ne}{x} Ly$\alpha$ line are
probably contaminated with Fe-L lines and are therefore considered less
reliable.}
\label{tab:lines}
\centering
\begin{tabular}{|llllrr|}
\hline
Line ID  & Pred.~$\lambda$ & Primary & Secondary & $\Delta$v$_{\rm Primary}$
& $\Delta$v$_{\rm Secondary}$  \\
&  \AA\  & photons (10$^{48}$s$^{-1}$) &  photons (10$^{48}$s$^{-1}$) & km s$^{-1}$ & km s$^{-1}$ \\
\hline
\ion{Si}{xiv} Ly$\alpha$ &  6.180 & 0.22$\pm$0.03 &  0.05$\pm$0.02 & $-$ 940$\pm$290 & + 470 $\pm$1650  \\
\ion{Mg}{xii} Ly$\alpha$ &  8.419 & 0.25$\pm$0.03 &  0.12$\pm$0.03 & $-$ 1000$\pm$360 & + 290 $\pm$1030 \\
\ion{Ne}{x}   Ly$\alpha$ & 12.132 & 0.83$\pm$0.11 &  0.50$\pm$0.08 & $-$ 550$\pm$150 & + 1450 $\pm$590 \\
\ion{O}{viii} Ly$\alpha$ & 18.969 &  2.95$\pm$0.27 &  1.40$\pm$0.18 &  $-$ 470$\pm$100 & + 430$\pm$190 \\
\ion{O}{vii} f  &          22.101 &  8.96$\pm$0.55 &  1.29$\pm$0.24  &  $-$ 250$\pm$20 & + 360$\pm$160 \\
\ion{N}{vii} Ly$\alpha$ & 24.779 &  3.69$\pm$0.94 &  0.80$\pm$0.16  &   + 15$\pm$360 &  \\
\ion{C}{vi} Ly$\alpha$ &  33.734 &  4.69$\pm$0.68 &  1.21$\pm$0.31  &  $-$ 300$\pm$120 & $-$ 860$\pm$330 \\
\ion{C}{v} f    &  41.472        &   6.44$\pm$1.99 &  1.01$\pm$0.40 &  $-$ 380$\pm$160  &  \\
\hline
\end{tabular}
\end{table*}

\begin{figure}
\resizebox{\hsize}{!}{\includegraphics[angle=-90]{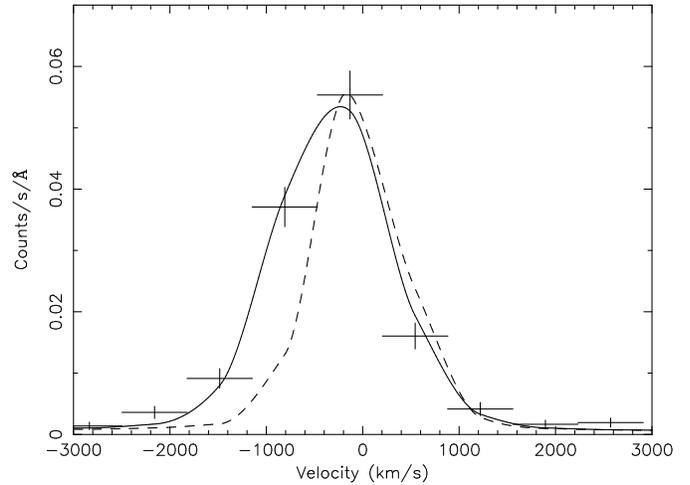}}
\caption{Spectrum of the \ion{O}{vii} forbidden line in $+1$ order for the Primary
region.  The velocity scale is set with respect to the rest frame of NGC~1068.
The solid line shows the best fit model with an average blueshift of $-250$~km/s
and a velocity broadening of $320$~km/s.  The dashed line is the best fit with no
velocity shift and broadening.  Note that in the latter case the line profile is
given by the zeroth order profile of the Primary as discussed in Sect.~3.1.
The bin size in wavelength space is $25$~m\AA.}
\label{fig:xx1}
\end{figure}

We have used a number of strong emission lines, the Ly$\alpha$ lines of Si, Al,
Mg, Ne, O, N and C, and the forbidden lines of the He-like triplets of O and N,
to look for velocity broadening (Table~\ref{tab:lines}.  All lines of the Primary region appear to be
broader than the zero order profile, although the formal errors are large.  The
most reliable velocity numbers are obtained from $\ion{O}{viii}$ Ly~$\alpha$ and
the forbidden line of the $\ion{O}{vii}$ triplet, $\sigma_{v}$ being 600 and 320
km/s, respectively.  The Si, Al, and Mg Ly$\alpha$ lines show values of about
1000~km/s but with large errors.  Fig.~\ref{fig:xx1} shows the \ion{O}{vii}
forbidden line of the Primary region in velocity space.

For the Secondary region the measured widths of the lines are consistent with
zero, with upper limits of about 1000 km~s$^{-1}$.

\begin{figure}
\resizebox{\hsize}{!}{\includegraphics[angle=-90]{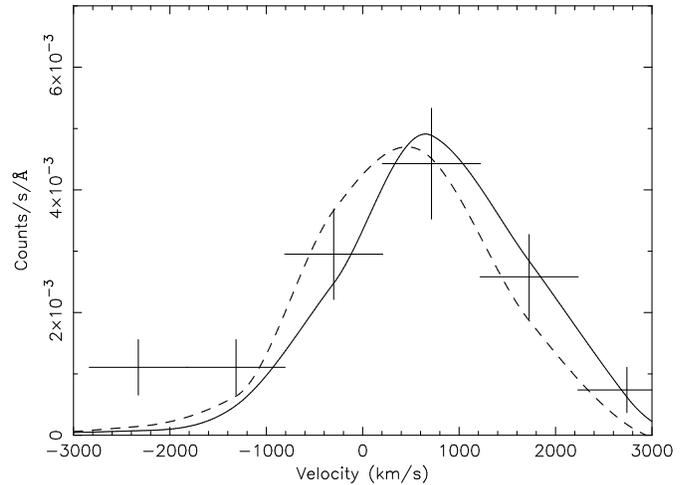}}
\caption{Spectrum of the \ion{O}{vii} forbidden line in $+1$ order for the
Secondary region.  The velocity scale is set with respect to the rest frame of
NGC~1068.  The solid line shows the best fit model with an average redshift of
$+360$~km/s and no velocity broadening.  The dashed line is the best fit with no
velocity shift and broadening.  Note that in the latter case the line profile is
given by the zeroth order profile of the Secondary as discussed in Sect.~3.1.
The bin size in wavelength space is $37.5$~m\AA.}
\label{fig:xx2}
\end{figure}

\subsection{Line Shifts}

We used the same set of lines to look for redshifts or blueshifts.  First of all
we find for most lines of the central region a difference between the measured
positive and negative spectral order position of about $0.02$~\AA.  One would
however expect a difference of $\sim 0.01$~\AA, based on the difference between
the peak position of the Gaussian fit to the zero order distribution and the
zero point of the wavelength scale.  The underlying assumption is that the line
radiation distribution coincides with the total zero order radiation
distribution.  We verified the line wavelength obtained above by fitting the
measured line shape to a Gaussian and forced the line width to be equal to the
zero order width; that has very little effect on the line wavelength
determination.  The results of fitting the positive and negative spectral order
lines together, are summarized in Table~\ref{tab:lines}.  We then looked for
possible differences in redshift or blueshift between the Secondary and Primary
region.

Fitting for the Secondary region, the positive and negative spectral orders
together lead to redshifts, also listed in Table~\ref{tab:lines}.  (All shifts
are calculated with respect to the systemic velocity).

\section{Long Wavelength L-shell Line Detections \label{sect:L-lines}}

We have also analysed the long wavelength part of the LETGS spectrum of the
Primary component (beyond 30~\AA).  This part of the spectrum contains the
L-shell complexes of Ne, Mg, Si, S, Ar and Ca.  We have determined the fluxes of
the strongest resonance lines in the $15-60$~\AA\ band (see
Table~\ref{tab:longwav}).  Wavelengths and oscillator strengths for these lines
are the same as those quoted in Kaastra et al.  (\cite{kaastra2002}).  The data
are taken from what is used in their {\it slab} and {\it xabs} models, and for
the relevant ions they are derived mostly from the compilation by Verner et al.
(\cite{verner96}); updates in some cases, using calculations with HULLAC
(Bar-Shalom et al.  \cite{bar-s}), were provided by Liedahl (see Kaastra et al.
\cite{kaastra2002} for more details).  In addition, we have included a number of
Fe-L transitions in the table.  For some ions we did not use the strongest line.
An example is \ion{Fe}{xix}, where the strongest lines are blended by lines from
\ion{Ne}{ix}.  Finally, in a few cases we co-added lines from the same ion if
they could not be resolved with the LETGS.

\begin{table}[!h]
\caption{Wavelengths, oscillator strength, ionization
parameters and line fluxes for
the strongest lines of each species in the Primary component.
Line strengths are given in units of $10^{48}$~ph/s.
}
\label{tab:longwav}
\centerline{
\begin{tabular}{|l|rrrrr|}
\hline
Ion & $\lambda$  & $f$ & $\log\xi$ & Flux       & Flux  \\
    & (\AA)      &   &           &  observed  & model \\
\hline
\ion{Ne}{viii} &57.75& 0.01& 1.0& 2.7$\pm$ 4.2& 2.2\\
\ion{Mg}{ix}   &48.34& 0.14& 1.2& 3.6$\pm$ 1.3& 3.1\\
\ion{Mg}{x}    &57.88& 0.22& 1.5& 6.0$\pm$ 5.0& 4.7\\
\ion{Si}{viii} &50.02& 0.12& 0.6& 2.9$\pm$ 1.4& 2.9\\
\ion{Si}{ix}   &55.30& 0.95& 0.9& 5.9$\pm$ 3.0& 4.5\\
\ion{Si}{x}    &50.52& 0.60& 1.3& 4.3$\pm$ 1.7& 5.1\\
\ion{Si}{xi}   &43.76& 0.48& 1.7& 4.7$\pm$ 1.0& 3.2\\
\ion{Si}{xii}  &40.91& 0.23& 2.0& 3.0$\pm$ 5.1& 2.2\\
\ion{S}{vii}  &51.81& 0.61& 0.0& 3.4$\pm$ 1.8& 3.9\\
\ion{S}{viii} &52.77& 0.89& 0.4& 3.7$\pm$ 2.3& 4.1\\
\ion{S}{ix}   &47.43& 0.38& 0.8& 2.7$\pm$ 1.0& 3.2\\
\ion{S}{x}    &42.54& 0.67& 1.1& 4.0$\pm$ 4.3& 3.3\\
\ion{S}{xi}   &39.24& 1.05& 1.4& 3.3$\pm$ 2.2& 4.1\\
\ion{S}{xii}  &36.40& 0.62& 1.7& 3.1$\pm$ 1.3& 2.5\\
\ion{S}{xiii} &32.24& 0.45& 2.0& 2.5$\pm$ 0.8& 2.1\\
\ion{S}{xiv}  &30.44& 0.34& 2.4& 1.6$\pm$ 1.0& 0.6\\
\ion{Ar}{x}    &37.43& 0.65& 1.0& 3.6$\pm$ 1.7& 1.4\\
\ion{Ar}{xi}   &34.33& 0.42& 1.3& 2.9$\pm$ 1.0& 1.0\\
\ion{Ar}{xii}  &31.39& 0.87& 1.5& 4.2$\pm$ 0.9& 1.5\\
\ion{Ar}{xiii} &29.32& 0.91& 1.8& 1.7$\pm$ 0.7& 1.2\\
\ion{Ar}{xiv}  &27.47& 0.65& 2.1& 0.7$\pm$ 0.5& 0.8\\
\ion{Ca}{xi}   &30.45& 2.34& 1.2& 1.6$\pm$ 1.0& 1.5\\
\ion{Ca}{xii}  &27.97& 1.30& 1.3& 0.2$\pm$ 0.4& 1.0\\
\ion{Ca}{xiii} &25.53& 0.47& 1.6& 0.3$\pm$ 0.2& 0.4\\
\ion{Ca}{xiv}  &24.13& 0.89& 1.9& 0.4$\pm$ 0.2& 0.6\\
\ion{Ca}{xv}   &22.73& 0.93& 2.1& 0.5$\pm$ 0.2& 0.4\\
\ion{Ca}{xvi}  &21.45& 0.66& 2.4& 0.9$\pm$ 0.5& 0.2\\
\ion{Ca}{xvii} &19.56& 0.65& 2.5& 0.2$\pm$ 0.1& 0.2\\
\ion{Ca}{xviii}&18.70& 0.37& 2.8& 0.3$\pm$ 0.3& 0.1\\
\ion{Fe}{xiv}  &58.96& 0.28& 1.4& 5.4$\pm$ 8.5& 4.5\\
\ion{Fe}{xv}   &52.91& 0.41& 1.6& 4.3$\pm$ 2.5& 4.7\\
\ion{Fe}{xvi}  &50.35& 0.16& 1.6& 5.4$\pm$ 1.9& 2.4\\
\ion{Fe}{xvii} &15.01& 2.72& 2.1& 1.6$\pm$ 0.3& 1.6\\
\ion{Fe}{xviii}&14.20& 0.88& 2.3& 0.8$\pm$ 0.2& 1.2\\
\ion{Fe}{xix}  &13.80& 0.21& 2.5& 0.4$\pm$ 0.2& 0.3\\
\ion{Fe}{xx}   &12.84& 0.93& 2.8& 0.7$\pm$ 0.2& 0.4\\
\ion{Fe}{xxi}  &12.29& 1.24& 3.0& 0.3$\pm$ 0.1& 0.4\\
\ion{Fe}{xxii} &11.71& 0.67& 3.1& 0.2$\pm$ 0.2& 0.3\\
\ion{Fe}{xxiii}& 8.30& 0.14& 3.3& 0.1$\pm$ 0.1& 0.1\\
\ion{Fe}{xxiv} &10.64& 0.37& 3.4& 0.2$\pm$ 0.2& 0.2\\
\hline
\end{tabular}
}
\end{table}

Given the limited and not homogeneously tested accuracy of the wavelengths, and
also the possibility of some blending we cannot always be sure of the derived
fluxes in each individual case.  However, we think that the set of fluxes still
reflects the overall behaviour of these ions.  Not all the fluxes are
statistically significant, but for the weakest lines the value given plus its
nominal error bar gives at least an order of magnitude estimate of the column
density of the parent ion.  Table~\ref{tab:longwav} lists the ionization
parameter, $\xi$, for which the ion concentration reaches a maximum.  We have
also made model predictions for the line fluxes.  We assume that for each line
the emitted flux is equal to the number of photons taken away from the power law
continuum ($\Gamma=1.7$, covering factor times luminosity $fL_X=10^{43}$~erg/s,
cf.  the RGS analysis of Kinkhabwala et al.  \cite{kinkhabwala02a}).  To be more
specific, we approximate the equivalent width for the absorption line by:
\begin{equation}
\label{eqn:w}
W = {\lambda \sigma_{\rm v}\over c} \int\limits_{-\infty}^{\infty}
[1-\exp (-\tau_0\mathrm{e}^{\displaystyle -y^2/2})]
{\mathrm d}y,
\end{equation}
with $\tau_0$ equal to the optical depth of the line at the line center,
given by
\begin{equation}
\label{eqn:tau}
\tau_0 = 0.106 f N_{16} \lambda / \sigma_{\rm v,100}.
\end{equation}
Here $f$ is the oscillator strength, $\lambda$ the wavelength in \AA,
$\sigma_{\rm v,100}$ the velocity dispersion in units of 100~km/s and $N_{16}$ the
column density of the ion in units of $10^{16}$~cm$^{-2}$.

In the above, we neglected the damping wings of the lines.  Equation \ref{eqn:w}
gives the equivalent width for an absorption line.  The power law continuum is
given by ${\mathrm d}N/{\mathrm d\lambda} = 8.67\times 10^{49} \lambda^{-0.3}$
with $\lambda$ in \AA.  The predicted emission line flux is then simply $F=W
{\mathrm d}N/{\mathrm d\lambda}$.  We have divided the 40 emission lines into 4
groups of ten, each spanning a range in ionization parameter.  For each group,
we have determined the velocity dispersion $\sigma_{\rm v}$ and total hydrogen
column density $N_{\rm H}$ that best describe the observed line fluxes (assuming
Solar abundances).  The results are summarized in Table~\ref{tab:sigcol}.

\begin{table}[!h]
\caption{
Velocity dispersion and column densities for different
ranges in $\log\xi$.
}
\label{tab:sigcol}
\centerline{
\begin{tabular}{|l|rrrr|}
\hline
$\log\xi$ & $\sigma_{\rm v}$ & range & $\log N_{\rm H}$ & range \\
\hline
0.0$-$1.2 & 170 & 120$-$220 & 21.9 & lower limit \\
1.3$-$1.6 & 220 & 130$-$300 & 21.9 & 21.8$-$22.1 \\
1.7$-$2.3 & 150 & 120$-$210 & 21.8 & 21.5$-$22.0 \\
2.4$-$3.4 &  40 &   30$-$60 & 22.0 & 21.5$-$22.7 \\
\hline
\end{tabular}
}
\end{table}

All hydrogen column densities are close to the value of $\log N_{\rm H} = 21.9$,
and we have adopted that value uniformly in the predicted line fluxes of
Table~\ref{tab:longwav}.  For the velocity dispersions in that table we have
adopted the values of Table~\ref{tab:sigcol} instead.  We find evidence for a
significantly lower velocity dispersion at higher values of the ionization
parameter.  This may explain partly the problem we had with the H-like Ne and Mg
lines, where we had fixed $\sigma_{\rm v}$ to $100$~km/s.  The tendency of a
smaller velocity dispersion for the most highly ionized material is also found
in the Seyfert~1 galaxy NGC~5548, where the high ionization line of
\ion{Si}{xiv} Ly$\alpha$ is consistent with a velocity component that in the
corresponding UV spectra shows up as much narrower than the lower ionized lines.

\section{Summary and Conclusions}\label{sec:discussion}

We have shown that the {\it Chandra} {\small LETGS} spectrum of the Primary
region (comprising the bulk of the flux from {\small NGC}~1068) is consistent
with emission from a photoionized and photoexcited cone of warm plasma, which is
in accord with previous observations of this object with {\it XMM-Newton}
{\small RGS} (Kinkhabwala et al.  \cite{kinkhabwala02b}).  We have also shown,
using the spatial resolution capability for angles of a few arcsec of the {\it
Chandra} grating spectrometers, that the X-ray spectrum of a separate region
(4$\arcsec$ to the {\small NE}), which we refer to as the Secondary, can be
explained by a similar model, except with a factor of a few lower inferred
radial ionic column densities.  Any additional emission in either spectrum due
to hot, collisional plasma is negligible.

The lower radial ionic column densities of the Secondary region as compared with
the Primary region are naturally explained in the context of an irradiated cone
of plasma.  The radial ionic column density through each region is given in
Kinkhabwala et al.  (\cite{kinkhabwala02b}) as:
\begin{eqnarray}
\lefteqn{N^{\mathrm{rad}}_{\mathrm{ion}}\simeq1.6\times10^{24}A_Z\frac{g}{fr_{\mathrm{min,pc}}}\times}\nonumber\\
&&\hspace{.6cm}\bigg[\frac{f_i}{0.5}\bigg]\bigg[\frac{(fL_X)}{10^{43}~\mathrm{ergs~s}^{-1}}\bigg]
\bigg[\frac{\xi}{1~\mathrm{ergs~cm~s}^{-1}}\bigg]^{-1}~\mathrm{cm}^{-2},
\label{eqn:N_ion}
\end{eqnarray}
where $A_Z$ is the abundance relative to hydrogen, $g$ is the radial filling
factor, $f=\Omega/4\pi$ is the usual covering factor, $r_{\mathrm{min,pc}}$ is
the minimum radius of the cone from the nucleus in pc, and $f_i$ is the
fractional ionic abundance.  The radial dependence of $r^{-1}$ in this equation
provides a simple explanation for the observed factor of a few drop in column
densities in the Secondary region versus those inferred for the Primary region
(Table~\ref{tab:densities}), which is closer to the nucleus by at least a
similar factor of a few.

An estimate of the radial filling factor of the plasma in the Secondary region
is also made possible by our knowledge of $r_{\mathrm{min}}$ ($\sim$4\arcsec;
$288$~pc) for this region.  The inferred radial column density of
$N^{\mathrm{rad}}_{\mathrm{ion}}=2.10^{17}$ for \ion{O}{vii} in the Secondary
region, along with $A_Z=4.89\times10^{-4}$ and $f<0.1$, imply a radial filling
factor of $g<0.07$ for plasma at the typical ionization parameter for
\ion{O}{vii} of $\xi=10$~ergs~cm~s$^{-1}$.

The remarkable overlap of the [\ion{O}{iii}] and X-ray cones appears to indicate
that these regions are related (Young et al.  \cite{young}).  X-ray spectroscopy
of the Primary region reveals excess blueshifts, whereas spectroscopy of the
Secondary region reveals excess redshifts.  These blueshifts and redshifts are
comparable to the shifts observed in the same respective regions for optical/UV
lines
(Cecil et al. \cite{cecil1990}, \cite{cecil2002};
Crenshaw \& Kraemer \cite{crenshaw}), kinematically linking the X-ray
emission regions with lower ionization state emission regions.  In addition,
Kraemer \& Crenshaw (\cite{kraemer}) argue that the optical/UV emission in the
cone is also ultimately powered by photoionization due to the inferred nuclear
continuum.

Using both K-shell and L-shell line series, we find that the relative abundances
of all observed elements are consistent with Solar abundances, aside from N.
From the relative strength of the column densities derived from K-shell emission
features, we find evidence for a relative factor of 2$-$3 more N compared with C
or O.  Further investigation of elemental abundances is left for the future.

The K-shell transitions as well as the L-shell transitions for multiple ions,
imply that a broad range in ionization parameter is present.  The long
wavelength emission line spectrum is very similar to what is found in Seyfert~1
galaxies in absorption.  These observations provide further confirmation of the
unification schemes, wherein Seyfert~1 and Seyfert~2 galaxies constitute the
same objects seen from different viewing angles.  The data also suggest a
variation of velocity dispersion as a function of ionization parameter (see
Table~\ref{tab:sigcol}).

\begin{acknowledgements}

{\small SRON}, The National Institute for Space Research, is supported
financially by {\small NWO}, the Netherlands Organization for Scientific
Research.  The Columbia University team is supported by {\small NASA}.
{\small AK} acknowledges additional support from an {\small NSF} Graduate
Research Fellowship and {\small NASA} {\small GSRP} fellowship.  {\small MS}
was partially supported by {\small NASA} through {\it Chandra} Postdoctoral
Fellowship Award Number {\small PF}01-20016 issued by the {\it Chandra} X-ray
Observatory Center, which is operated by the Smithsonian Astrophysical
Observatory for and behalf of {\small NASA} under contract {\small NAS}8-39073.
\end{acknowledgements}

\end{document}